\documentclass{emulateapj}

\usepackage{subfigure}
 \usepackage{graphicx}
\usepackage{threeparttable}
\usepackage{rotating}
\usepackage{lipsum}
\usepackage{arydshln}
\usepackage{amsmath}
\usepackage{hyperref}
\usepackage{booktabs}
\usepackage{xcolor}
\usepackage{soul}

\newcommand{\lya}{Lyman-$\alpha$}
\newcommand{\lum}{$L_{Ly\alpha}$}

\newcommand{\wlya}{\hbox{$W_{\mathrm{Ly}\alpha}$}}
 
 \newcommand{\ergscm}{$\rm\;erg\;s^{-1}cm^{-2}$}
 
\newcommand{\ergscma}{$\rm\;erg\;s^{-1}\;cm^{-2}\;\AA^{-1}$}
\newcommand{\ergs}{$\rm\;erg\;s^{-1}$}

\newcommand{\figsid}{$\rm\;FIGS\_GN1\_1292 $}

\newcommand{\dtoprule}{\specialrule{0.8pt}{0pt}{0.4pt}%
            \specialrule{0.2pt}{0.4pt}{\belowrulesep}%
            }
\newcommand{\dbottomrule}{\specialrule{0.pt}{0pt}{0.pt}%
            \specialrule{0.2pt}{0pt}{\belowrulesep}%
            }            
\newcommand{\paa}{PA=--56}
\newcommand{\pab}{PA=--98}

\newcommand{\ed}[1]{{{#1}}}

\slugcomment{}

\shorttitle{First Results From FIGS Survey}
\shortauthors{Tilvi et al.}

\begin{document}

\title{First Results From Faint Infrared Grism Survey (FIGS): First Simultaneous Detection of Lyman-$\alpha$ Emission and Lyman Break from  a  galaxy at \it{z}=7.51
}


\author{V. Tilvi \altaffilmark{1},
N. Pirzkal\altaffilmark{2},
S. Malhotra\altaffilmark{1},
S. L. Finkelstein \altaffilmark{3},
J. E. Rhoads  \altaffilmark{1}, 
R. Windhorst  \altaffilmark{1},
N. A. Grogin  \altaffilmark{2},
A. Koekemoer  \altaffilmark{2},
N. L. Zakamska   \altaffilmark{4,5},
R. Ryan \altaffilmark{2},
L. Christensen  \altaffilmark{6},
N. Hathi  \altaffilmark{7}, 
J. Pharo \altaffilmark{1},
B. Joshi \altaffilmark{1},
H. Yang \altaffilmark{1},
C. Gronwall   \altaffilmark{8,9},
A. Cimatti  \altaffilmark{10},
J. Walsh  \altaffilmark{11},
R. O'Connell  \altaffilmark{11},
A. Straughn  \altaffilmark{13},
G. Ostlin  \altaffilmark{14},
B. Rothberg  \altaffilmark{15},
R. C. Livermore  \altaffilmark{3},
P. Hibon  \altaffilmark{16}, and 
Jonathan  P. Gardner  \altaffilmark{13}
}

\altaffiltext{1}{School of Earth \& Space Exploration, Arizona State University, Tempe, AZ. 85287, USA}
\altaffiltext{2}{Space Telescope Science Institute, Baltimore, MD. 21218, USA}  
\altaffiltext{3}{Department of Astronomy, The University of Texas at Austin, Austin, TX. 78712, USA}  
\altaffiltext{4}{Deborah Lunder and Alan Ezekowitz Founders' Circle Member, Institute for Advanced Study, Einstein Dr., Princeton, NJ 08540, USA}
\altaffiltext{5}{Department of Physics \& Astronomy, Johns Hopkins University, Bloomberg Center, 3400 N. Charles St., Baltimore, MD 21218, USA}
\altaffiltext{6}{1 Dark Cosmology Centre, Niels Bohr Institute, University of Copenhagen, Juliane Maries Vej 30, 2100 Copenhagen, Denmark  }
\altaffiltext{7}{Aix Marseille Université, CNRS, LAM (Laboratoire d\'Astrophysique de Marseille) UMR 7326, 13388, Marseille, France}
\altaffiltext{8}{Department of Astronomy and Astrophysics, The Pennsylvania State University, University Park, PA 16802, USA}
\altaffiltext{9}{Institute for Gravitation and the Cosmos, The Pennsylvania State University, University Park, PA 16802, USA}
\altaffiltext{10}{Dipartimento di Fisica e Astronomia, Universit\'a di Bologna, Alma Mater Studiorum, viale Berti Pichat 6/2, I-40127 Bologna, Italy }
\altaffiltext{11}{European Southern Observatory, Karl-Schwarzschild Strasse 2, D-85748 Garching, Germany }
\altaffiltext{12}{Department of Astronomy, University of Virginia, Charlottesville, VA 22904-4325, USA}
\altaffiltext{13}{Astrophysics Science Division, Goddard Space Flight Center, Code 665, Greenbelt, MD 20771, USA }
\altaffiltext{14}{Department of Astronomy, Stockholm University, Oscar Klein Center, AlbaNova, Stockholm SE-106 91, Sweden }
\altaffiltext{15}{Large Binocular Observatory, Tucson, AZ 85721, USA}
\altaffiltext{16}{Gemini South Observatory, Casilla 603, La Serena, Chile}

\begin{abstract} 
Galaxies at high redshifts provide a valuable tool to study \textit{cosmic dawn}, and therefore it is crucial 
to reliably identify these galaxies.
Here, we present an unambiguous  and first simultaneous detection of both the \lya\ emission  and the  Lyman break from a
$z=7.512\;\pm\;0.004$ galaxy, observed in the Faint Infrared Grism Survey (FIGS). 
These spectra, taken with G102 grism on Hubble Space Telescope (HST),  show a
significant emission line detection ($6\sigma$)  in \ed{two} observational position
angles (PA), 
with 
\lya\ line flux  of 
$1.06\pm 0.19 \times10^{-17}$\ergscm.
The line flux is nearly a factor of four  higher than 
\ed{in  the archival }
 MOSFIRE spectroscopic observations.
This is consistent with other recent observations implying that ground-based near-infrared spectroscopy 
underestimates total emission line fluxes, and if confirmed,  can have  strong implications
for reionization studies that are based on ground-based \lya\  measurements.
A 4-$\sigma$ detection of the NV line in  one PA also
suggests a weak Active Galactic Nucleus (AGN), 
\ed{and if confirmed would make} 
this source the highest-redshift AGN yet found. 
These  observations from the {\it Hubble Space Telescope} thus 
clearly demonstrate the sensitivity of the FIGS survey, and the capability of grism
spectroscopy to study the epoch of reionization.
  \end{abstract}

\section{Introduction} 
To gain a complete understanding of the early universe, it is crucial to reliably identify
high-redshift galaxies, because   the formation and evolution of the earliest galaxies and the ionization 
of the  intergalactic medium (IGM) during the epoch of reionization are deeply intertwined.
The current consensus is   that the IGM is mostly ionized   at 
$z<7\;$\citep{fan02,mr04}, and  therefore, the process of reionization must have occurred 
at $z\gtrsim7$, where the IGM  is expected to be significantly neutral. 
This is also consistent with recent Planck results\;\citep{pla15},
which find an electron scattering optical depth
equivalent to  
an instantaneous reionization event  at $z\approx\;8.8$.

\lya\ emission from star-forming galaxies provides a unique probe of reionization. 
This is because \lya\ flux is attenuated by the  neutral hydrogen in
the IGM as  \lya\ photons are resonantly scattered out of
the line of sight when passing through a neutral IGM. This should result in a decrease in \lya\
emitting galaxy counts at $z>6$\;\citep{rho01, hu02, mr04}.
In addition to being a probe of reionization,
galaxies in the early universe likely contributed significantly to the reionization process\;\citep{oes10, mcl10, fin12, rob13, fin15, bou15a}.
 Furthermore, the $z>6$ universe provides the best chance to 
discover pristine galaxies\;\citep{sob15}, and therefore identification of high-$z$ galaxies is essential to gain a critical
insight into the   early universe.

In recent years, significant progress has been made in identifying hundreds of galaxy candidates at
$z>7$, \ed{using extremely deep imaging observations from the {\it Hubble Space Telescope}
\citep[HST; e.g.,][and references therein]{fin15, bou15a}.}
These galaxy candidates,  referred to as `Lyman-break'  galaxies (LBGs),   are 
primarily selected based on the \lya\ break at 1216\;\AA\ caused by the intervening 
neutral hydrogen in the IGM.
While there have been several spectroscopic followup observations of $z>7$ LBGs, only a handful of galaxies
have yielded  spectroscopic redshifts via detection of either  the \lya\ emission line\;\citep[e.g.,][]{van11, ono12,  shi12, fin13, sch14, oes15, zit15}
or the Lyman-break\;\citep{wat15,oes16}.

Galaxy searches at $z>7$ have also  been carried out 
using a narrow-band imaging technique in 
which galaxies are preselected to have a strong \lya\ line (known as \lya\ emitting galaxies; LAEs).
This technique has been successfully employed to
identify many LAE candidates out to $z>7.5$\;\citep[e.g.,][]{hib10,til10,cle12,kru12}

Despite spectroscopic successes 
at $z<7$, spectroscopic confirmations of a large sample of galaxies 
at $z>7$ has
been challenging.  
Recent studies, based on  spectroscopic observations of $z>7$ galaxies,  
 have claimed a precipitous drop in the observed  number of \lya\ emitting galaxies among
LBGs\;\citep{car12,  tre13, til14,  fai14,  sch14}.
However, 
it is not obvious whether the dominant factor behind the non-detection
of expected \lya\ emission is due to small-number statistics, evolving galaxy properties, 
an observational selection bias, or increasing neutral hydrogen at $z>7$.
These issues are further complicated by  the presence of abundant  atmospheric 
night-sky lines at near-infrared wavelengths, contaminating emission
lines in the ground-based spectra.

Fortunately, many of the  above issues can be circumvented using space-based 
slitless grism spectroscopy\;\citep[e.g.,][]{mal05, bra12,  van13,sch14b, tre15}
 because it eliminates
the near-infrared atmospheric contamination.
Recently,\;\citet{sch16} have found several $z>7$ candidates using the grism data
obtained from the 
Grism Lens-Amplified Survey from Space\;\citep{tre15}.
Furthermore, 
 spectroscopic
redshifts have been measured using a continuum detection of the Lyman-break\;\citep{rho13, oes16}
even in the absence of a \lya\ line.
This is critical because while there are other  emission lines available for measuring redshifts
\ed{\citep[see e.g., ][and references there in]{sta16}},
 they are much weaker compared to
the \lya\ line, and therefore, continuum Lyman-break detection provides a promising tool 
to 
measure spectroscopic redshifts during the epoch of reionization.

Here we present the first results from the Faint Infrared Grism Survey
(FIGS; Malhotra et al.  [in prep]), currently  the most sensitive G102 grism survey.
In this paper, we present the G102 slitless grism spectroscopic observations of \figsid, 
a $z=7.51$ galaxy
in the GOODS-N\;(GN1) field, 
which has a  ground-based spectroscopic redshift based on the   \lya\ emission line detection.
Using the  
HST grism observations, this is now the highest redshift galaxy that has simultaneous detection of 
Ly$\alpha$ emission line  and continuum Lyman-break. 
 In \S2 we present our observations  and 
spectral extraction.   In \S3 we present our results, and 
compare our observations to those from the ground, and in \S4 we
summarize our conclusions.
 Throughout this paper, we use AB magnitudes, and 
$\Lambda$-CDM cosmology
with $H_0=70.0$ km s$^{-1}$Mpc$^{-1}$, $\Omega_m$=0.27, and $\Omega_\lambda=0.73$.

\section{Observations: FIGS survey}

The FIGS survey is a 160-orbit  G102 WFC3/grism survey, designed to obtain
40-orbit depth spectroscopic  observations in two GOODS-N and two GOODS-S fields.
To minimize 
contamination 
\ed{by} overlapping 
spectra from nearby objects, each 
field was observed at five different position angles (PA; each at an 8-orbit depth).
\ed{ The 5 PA survey strategy was found to be optimal in minimizing 
 the contamination in the spectra of  $z >6$ sources based on aXesim simulations (see Malhotra et al in preparation).
 }
\ed{The total exposure time for the 40-orbit GN1 field is 101,100 sec; for the 5 individual PAs the exposure
 time varied from 19,300 to 21,900 sec.}
For complete details about the FIGS survey, we refer the reader to Malhotra et al (in preparation).

\begin{figure*}[t!]
  \centering
  \begin{minipage}[b]{0.45\textwidth}
    \includegraphics[width=\textwidth]{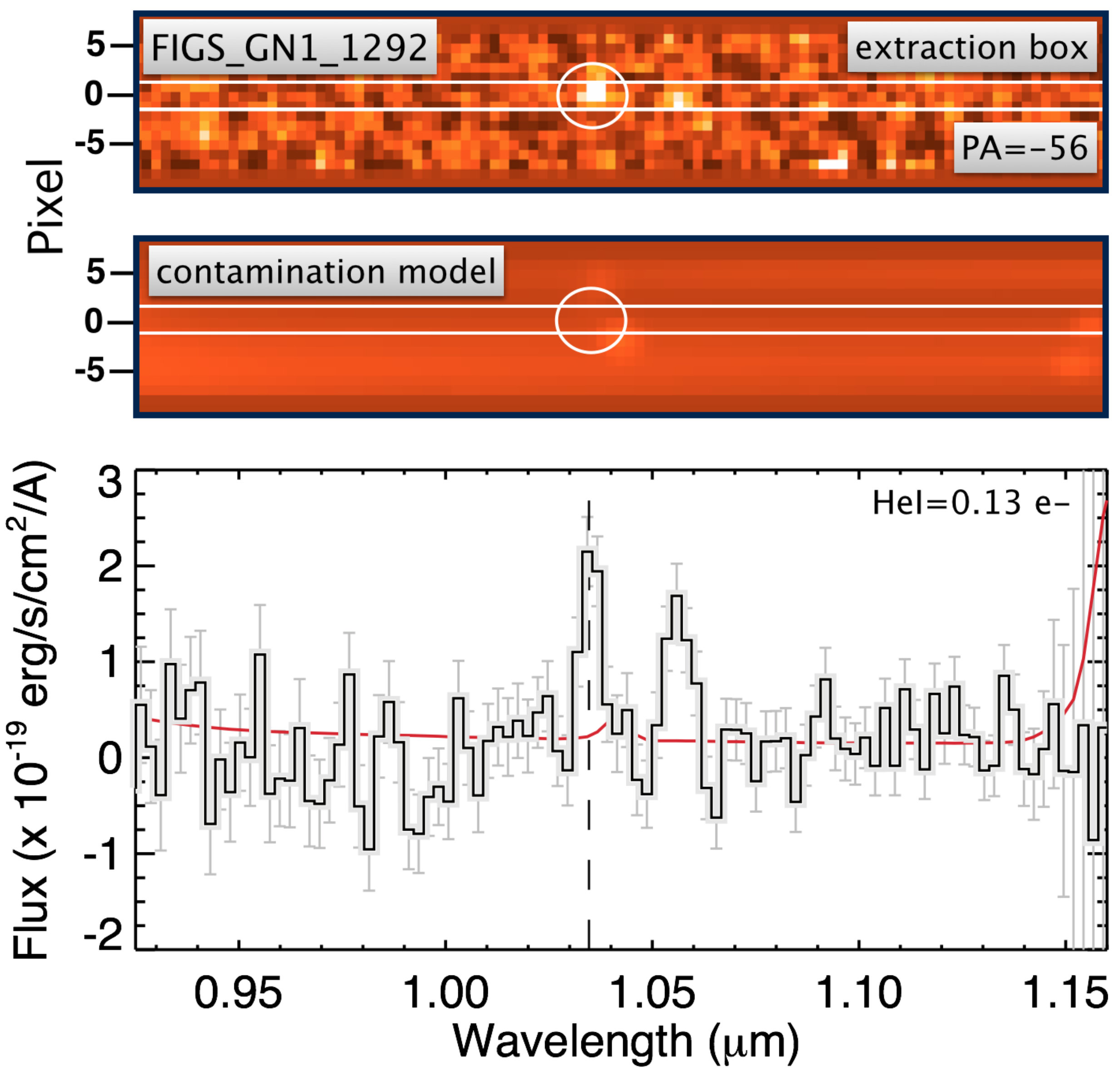}
  \end{minipage}
  \hfill
  \begin{minipage}[b]{0.45\textwidth}
    \includegraphics[width=\textwidth]{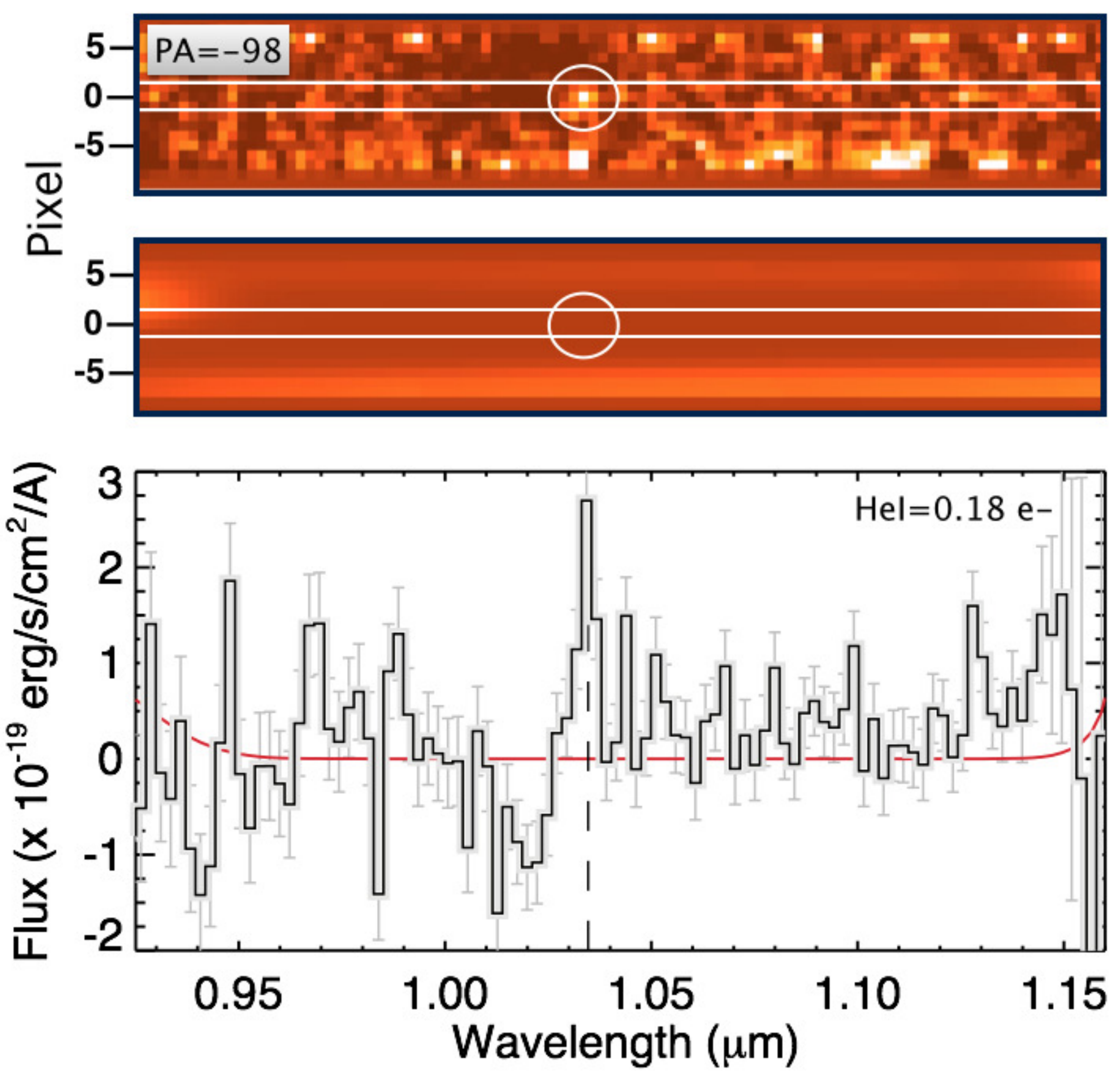}
  \end{minipage}
      \caption{
      Spectra of \figsid\ in two PAs used in this work. 
      Left: The top panel shows the 2D spectrum in  
        PA=$-$56 (not corrected for contamination).  The middle panel
      shows the contamination model,  while the bottom panel shows the 
      contamination-corrected 1D spectrum. The red line 
      in the bottom panel shows the contamination level.
      The rectangular region in the 2D spectrum  is the extraction
      area (3 pixels wide), and the circle represents the detected
      emission line. 
       A second possible emission feature is seen near 
      $1.055\mu$m in this PA.  Possible interpretations of this feature
      are discussed in section~\ref{sec:NV}.
      The HeI sky-background (noted in legend) significantly varies among different
      PAs, decreasing the sensitivity of spectra (see \$3.1 for details).
      Right: Same as left panels but for the PA=-98 observations.}
  \vspace{0.5cm}
\end{figure*}

\subsection{2-Dimensional Spectral Extraction}

We used the grism extraction software package aXe\footnote{http://axe-info.stsci.edu/}
(Pirzkal et al. 2001, K\"ummel et al. 2009) 
to extract individual sources. The method is similar to that of  the GRism 
ACS Program for Extragalactic Surveys (GRAPES: Pirzkal et al. 2004), but includes additional steps
 necessary to handle the HST WFC3 infrared data of FIGS. 
 
 First, 
 a master 
catalog of sources was generated using deep mosaics from the CANDELS survey
\citep{koe11,gro11} in the z, J, and 
H bands. These mosaics also served as our absolute astrometric reference points for the FIGS 
F105W direct images and the associated dither G102 exposures. They also allowed us to include
the colors of sources when computing spectral contamination. Special care was taken to subtract 
the varying backgrounds  from the  individual WFC3 exposures. This includes the HeI varying background 
(Brammer et al 2014\footnote{ISR WFC3 2014-03}, Sabbi et al 2015\footnote{WFC3 ISR 2015-07})
and 
the zodiacal light background levels, 
 and allowed for the
use of up-the-ramp fitting to remove cosmic rays from individual WFC3 exposures. 2-dimensional 
spectra were then extracted and combined at each position angle  using the aXeDrizzle
feature of aXe, which also removes the sky background.  
The end product is a set of multi-extension FITS files that each contain the 
spectrum of the science object (drizzled to the native pixel scale of 0\arcsec.128 per pixel, 
with a linearized wavelength scale),   an error estimate, a spectral
contamination model, and an effective exposure map. Details of the FIGS pipeline are given in
Pirzkal et al. (in preparation).

\vspace{0.1cm}

\subsection{1D Extraction}
We extracted 1D spectra of \figsid\ from the 2D multi-extension FITS file 
 by summing all pixels in the spatial direction (3 pixels wide) and collapsing it
 to a  single pixel at each wavelength.
 \ed{Based on our 3 and 5 pixel extraction widths, we found that  3-pixel extraction width  yields maximum 
 signal-to-noise (S/N) ratio for the \lya\ line, and therefore, in this study we use 3-pixel (0.384\arcsec) extraction width,
 which is also 
\ed{well matched to} the FWHM (0.36\arcsec) of \figsid.
 }
 To convert 1D spectra from  counts/sec 
 to physical units (\ergscma), 
 we used the following conversion:  flux [\ergscma] = flux [counts/sec]/sensitivity/dispersion, 
 where the sensitivity comes from the sensitivity function
 \footnote{WFC3.IR.G102.1st.sens.1.fits}
provided for the WFC3 grism, and the dispersion is the wavelength dispersion at each wavelength.
\ed{
Our preliminary estimates of the survey depth, based on inserting and recovering simulated sources,
 reaches the expected line flux limit of 
 $9\times\;10^{-18}$\;\ergscm\;(3$\sigma$) for a single PA.}

\section{Results and Discussion}
\subsection{Line Detection}
As can be seen in Figure 1, the emission line is clearly visible in both the  2D and 1D spectra  in 
two PAs.
This object
has a previous ground-based Keck/MOSFIRE
spectroscopic redshift of $z =\;7.5078\pm0.0004$, based on a faint \lya\ line detection
\citep{fin13}.

 \begin{figure}[t]
\epsscale{0.95}
\plotone{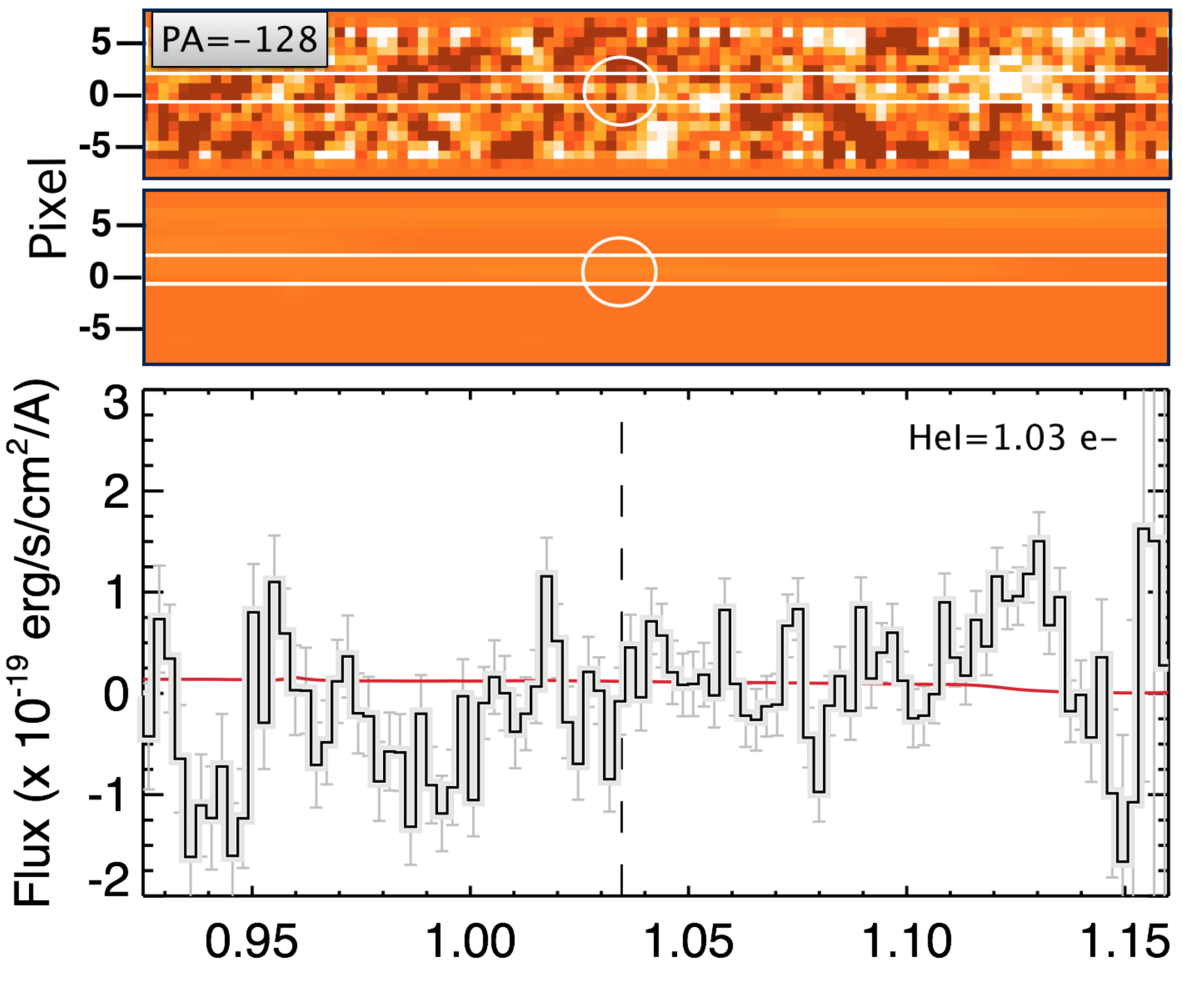}
\epsscale{0.95}
\plotone{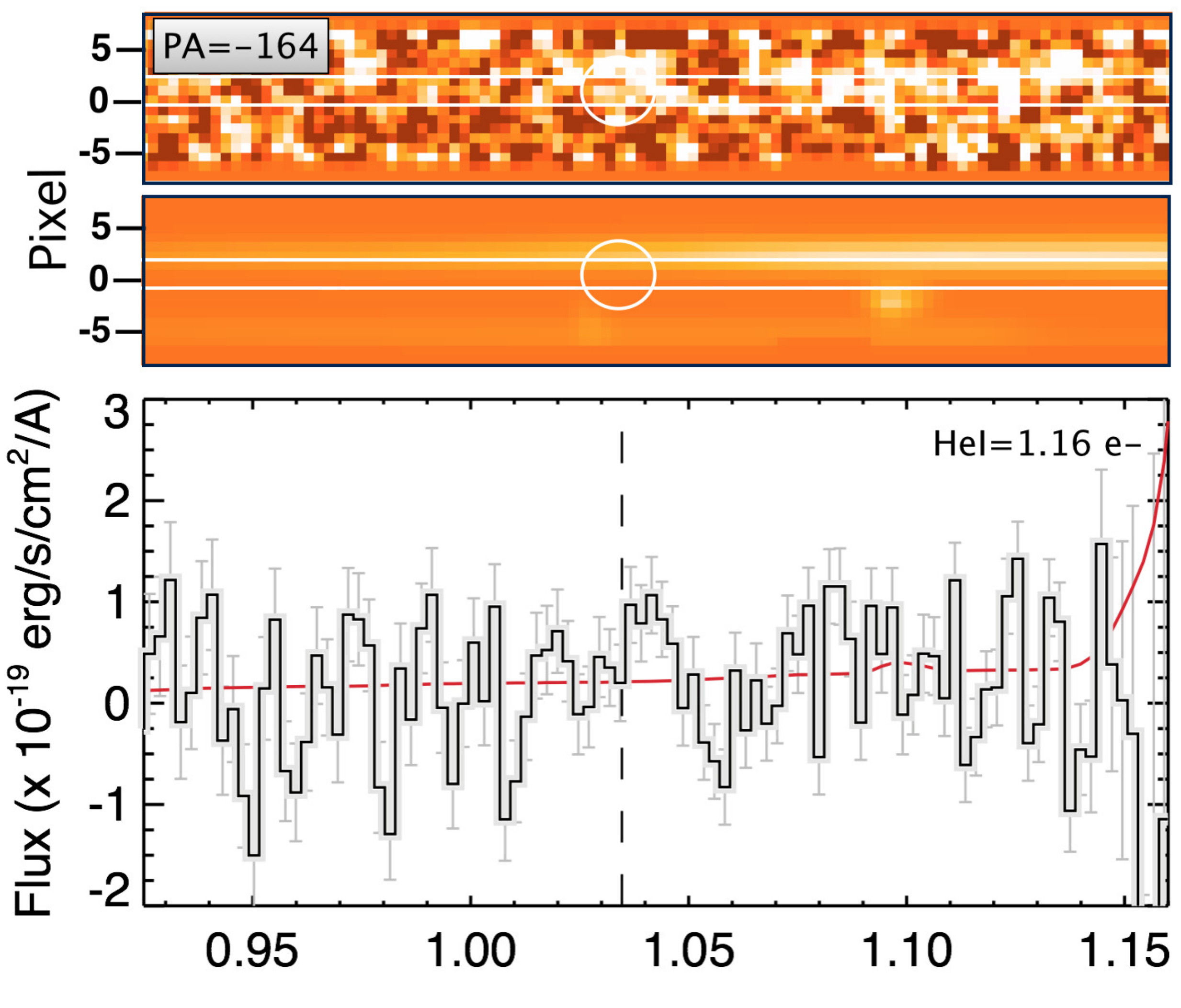}
\epsscale{0.95}
\plotone{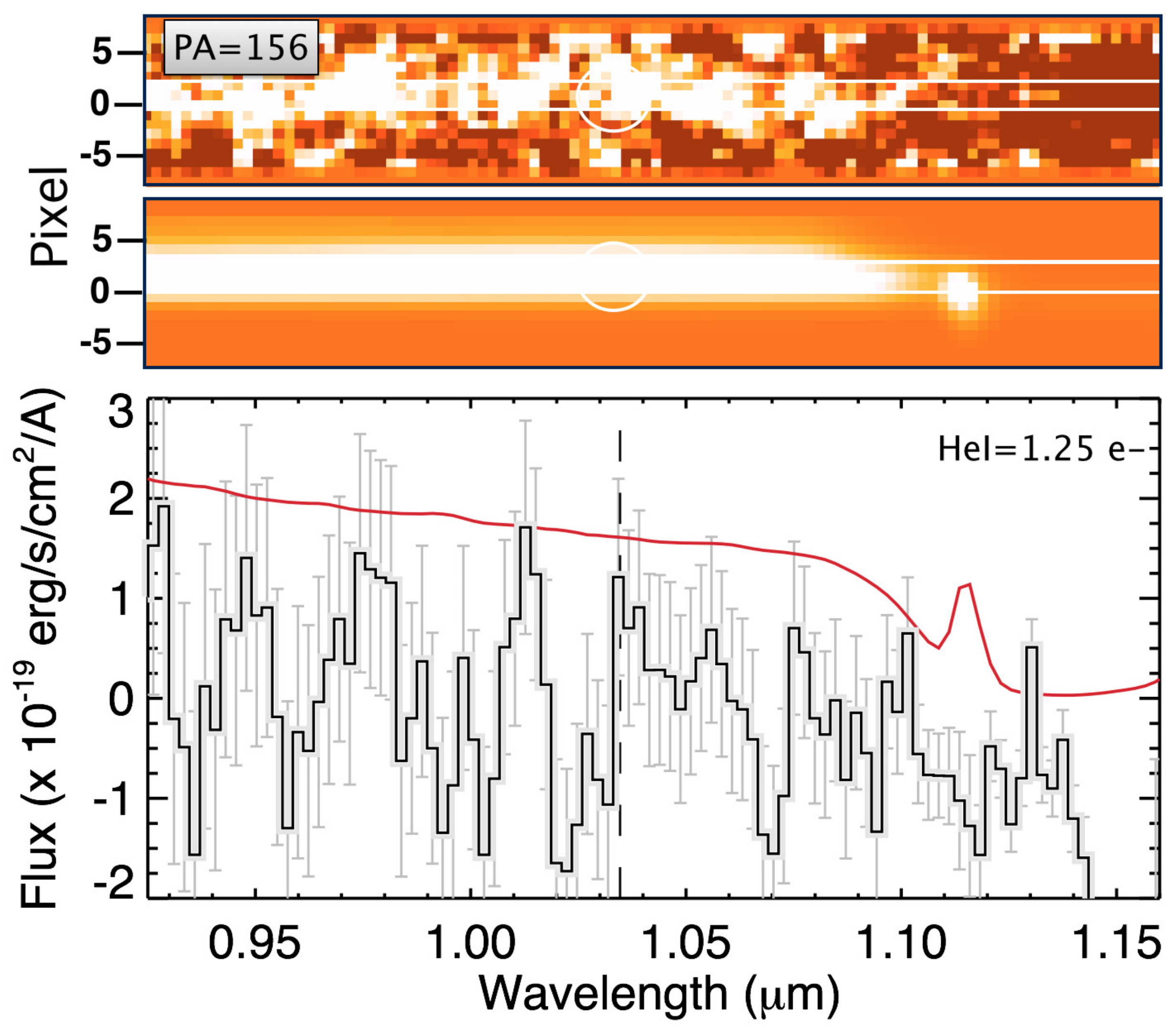}
\caption{
Same as Figure 1 except for the remaining three PAs where the \lya\ line is not detected. 
The vertical dashed line shows the position of the expected \lya\ line.
\vspace{0.3cm}
}
\end{figure}

For this galaxy,  while there are  five  different PAs  available  from
the FIGS data, three of the PAs (Figure 2) are significantly 
affected by varying HeI background (noted in legends). 
The three PAs where the line is not detected have HeI backgrounds that are
nearly 10$\times$ higher.
  \ed{
  This elevated HeI background\footnote{It is possible that our background noise is underestimated, however 
this does not change our conclusions in this paper.}
 does not vary significantly during the FIGS grism integrations, thanks to our
  survey design where we obtained direct imaging observations at the beginning or end of each orbit
  when the HeI background was expected to be highest and most variable.
  Thus we do not gain in S/N by discarding individual
  grism readouts.
  }
Therefore, to increase the signal-to-noise ratio in the stacked  spectrum (\S 3.3), we use data from only two PAs in this 
study.

\subsection{Possible Detection of NV}\label{sec:NV}
In  \paa, in addition to the \lya\  line at $\lambda\sim1.03\mu$m, there is another significant line 
\ed{($f_{\lambda} = 0.91\;\pm\;$0.21$\times$\;10$^{-17}$\;\ergscm)}
at  
$\lambda\sim1.055\mu$m, with a spatial offset of $\sim$0\arcsec.1, perpendicular to the dispersion direction.
This line, however, is not detected in \pab. 
Our careful inspection of 2D spectrum in \paa\ did not yield any $\rm\;0^{th}$,  $\rm\;1^{st}$, or 
2$^{nd}$ order contamination from other sources; our contamination models already include contamination
from these orders.
\ed{Furthermore, inspection of 1D spectra of contaminating sources do not show any strong emission line that could potentially
produce the tentative NV line at $\lambda\sim1.055\mu$m in PA=-56.}

Inspection of the ground-based spectrum 
found a marginal detection (2.5$\sigma$) at $\lambda\sim1.055\mu$m, however with a slightly larger  spatial offset from the \lya\ axis.
Thus, based on the offsets seen in the 2D grism and MOSFIRE spectra and the RGB image 
(Figure 3),  it is possible that this line is NV($\lambda\;1240$),
\ed{ a high ionization line, and 
}
 a signature of an (weak) 
Active Galactic Nucleus  
\ed{\citep[AGN; e.g., ][]{ham99}},
off-centered from the  \lya\ emitting region.
\ed{Furthermore, it is also  possible that  the NV emission is enhanced via resonant scattering of \lya\ photons   \ed{\citep[e.g.,][]{wan10}}.
}
Such off-axis  emission 
 from AGN reflection clouds has been seen before at lower redshifts \citep[][]{win98}, and
 it is argued that such off-axis emission could be missed in spectroscopy due to different position angles in both ground-based and space-based
spectroscopy.

\ed{In addition, we performed simulations by inserting and recovering artificial lines in the 
2D spectra, and  found that about 5\% of the times, a line as bright as the tentative NV
 line will remain undetected at $< 1\sigma$  significance.
Stronger conclusions will likely depend on deeper G141 grism data in the GN1 field.}
However, 
if the observed NV line is real, that would make this the highest-redshift AGN, and would support the idea that the (weak) AGNs   might help clear
 the surrounding neutral hydrogen
around  galaxies during  the epoch of reionization, making them visible in the \lya\  emission.

 \begin{figure}[t]
\epsscale{1.14}
\plotone{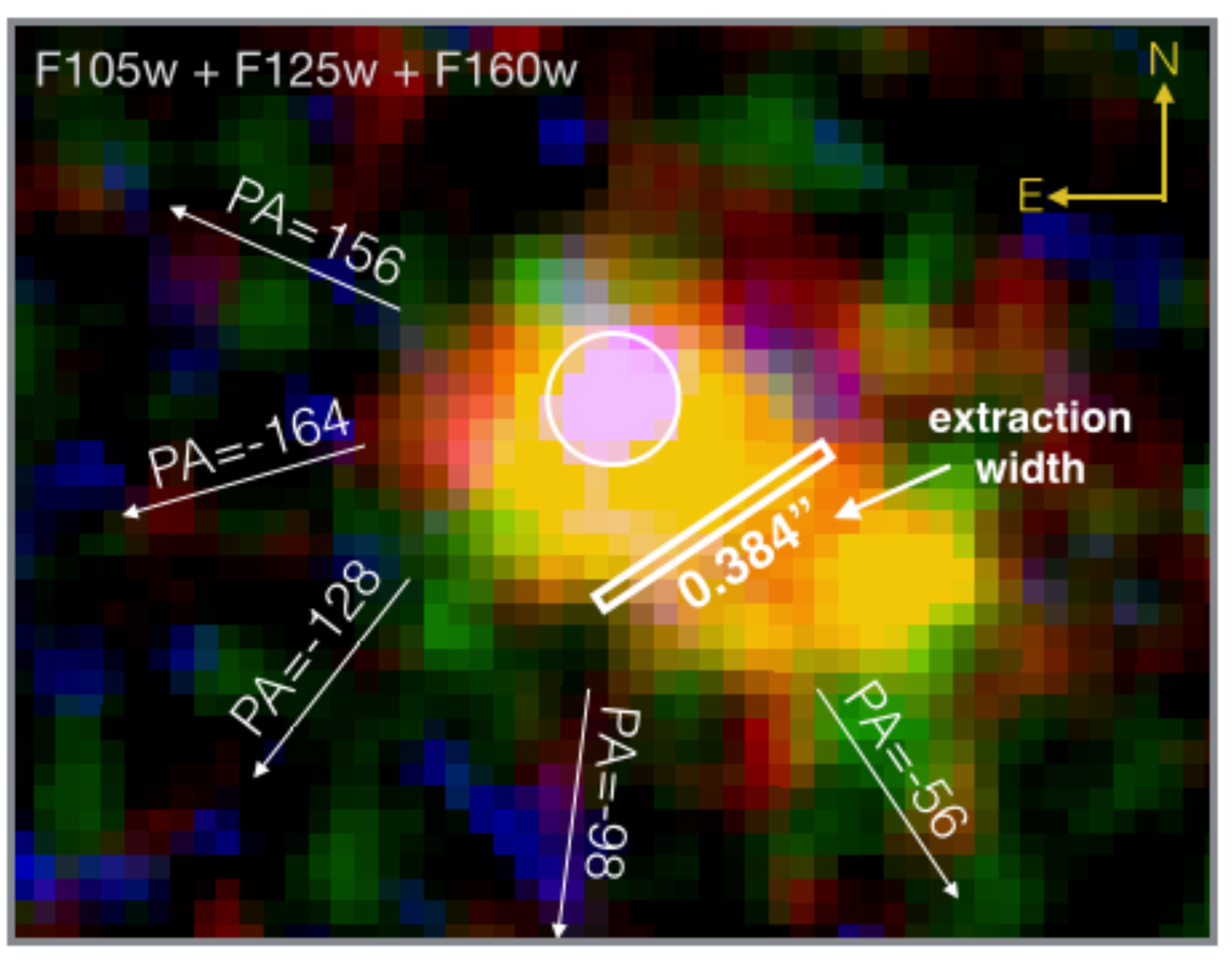}
\caption{
RGB composite of \figsid\ using HST/WFC3 broadband images. The off-centered white region enclosed within the
white circle is likely the source of the \lya\ emission.
The white rectangular box shows the spectral extraction width  (0\arcsec.384).
 In the 2D spectrum (Figure 1; left panel) there is a possible 
detection of NV. If real, this is likely coming from  a weak AGN at the  center of the galaxy.
\vspace{0.3cm}
}
\end{figure}

 \begin{figure}[t]
\epsscale{1.15}
\plotone{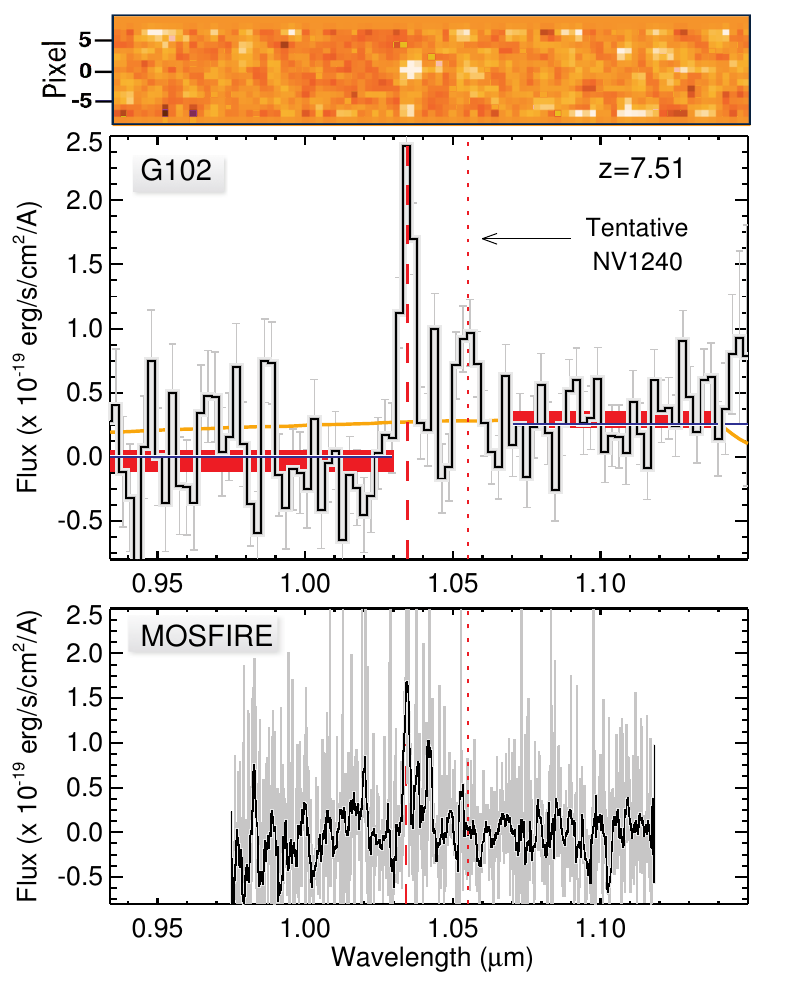}
\caption{\label{fig:stack}
Top: the average of two contamination-free PAs from Figure 1 for \figsid\ 
 in the GN1 field.  The middle panel shows 1D contamination-corrected spectrum, extracted using 3 pixels-wide aperture.
 As can be seen,   the \lya\ emission line is clearly visible in both
 2D and 1D spectra,  with wavelength 
$\lambda=1.0347\;\pm\;0.005\;\mu$m, consistent with the  ground-based Keck/MOSFIRE
spectroscopic detection (bottom panel) from Finkelstein et al.\
(2013).  \ed{ The orange horizontal line shows the sensitivity of G102, normalized to the redder continuum.}
The MOSFIRE spectrum shown in gray represents native resolution while black line shows a heavily smoothed
spectrum.
The FIGS HST grism spectrum also shows a clear detection of the continuum \ed{ (horizontal blue line, and 
 errors from bootstrap technique shown in shaded red region; see \S3.5)} with 5.6$\sigma$ significance measured 
at $\lambda=1.07-1.14\;\mu m$.
\vspace{0.3cm}
}
\end{figure}

\subsection{Emission Line Properties}
To increase the signal-to-noise ratio  of \figsid\ spectra, we combined the two PAs  (from Figure 1) 
to get average 2D and 1D spectra (Figure 4).
As can be seen the emission line appears well-detected in both 2D and 1D, and the emission line 
 wavelength (shown with vertical dashed line) matches
very well with the ground-based spectrum.

\ed{Alternative explanations for the $\lambda=1.0347\;\mu$m line are
  strongly disfavored.  H$\alpha$ is ruled out by the spectral break
  in the FIGS spectrum while  the [OIII] doublet is ruled out by the break
  {\it and} by the 
  absence of a [OIII]$\lambda$4959 line in the MOSFIRE spectrum
  \citep{fin13}.  
  The [OII]$\lambda$3727 doublet is hardest to rule out.  Line
  asymmetry would be useful in principle, but the MOSFIRE spectrum
  overlaps a night sky line, precluding reliable asymmetry
  measurement \citep{fin13}, and the FIGS data lack the needed spectral
  resolution.
  We  highly disfavor the line being [OII] emission due to non-detection of the
  [OIII] emission line at $\lambda\sim1.39\mu$m 
  in the archival  G141 grism data, and therefore 
  favor this line being \lya\ emission  from a $z=7.51$  galaxy.
}

To measure the  grism line properties we used a  Gaussian fitting function 
(MPFIT function in IDL)
to the
1D spectrum, shown in Figure 4 (middle panel).
We measured  the  \lya\ equivalent width (\wlya; the ratio of emission line flux
to the continuum flux density) 
using  the average continuum flux density between $\lambda=1.07\;\mu$m  and  
$\lambda=1.14\;\mu$m (Figure 4), which 
yields $f_{\lambda} = 2.52\;\pm\;$0.59$\times$\;10$^{-20}$\;\ergscma.
Combining this measurement with the \lya\ line flux, we get rest-frame \wlya\;=\;49.3\;$\pm\;8.9$\AA.  
 Other physical properties  are listed in Table 1.

\subsection{Comparison of Keck/MOSFIRE Spectroscopic and {\it HST} Grism
  Spectra}

The bottom panel in  Figure 4 shows the Keck/MOSFIRE ground-based spectrum of this source. 
The MOSFIRE spectrum (shown in gray) is contaminated by several
OH sky-line residuals, some of which are  even brighter than  the \lya\ line
itself,  making it difficult to observe in 
the near-infrared part of the spectrum from the  ground. 
The emission line at $\lambda=1.0343\;\mu m$ (marked by the vertical dashed line) is partly contaminated
by a night-sky line. The space-based grism observations do not suffer from this issue. 
On the other hand, the MOSFIRE spectrum has a much higher spectral
resolution which would potentially allow  to distinguish  between two 
closely spaced lines, as well as the shape of the \lya\ line that tends to be asymmetric at high redshifts.
With the G102 grism resolution, we cannot measure the shape of the \lya\ line.

The emission line wavelength in the \figsid\ spectrum matches very well with the
ground-based Keck/MOSFIRE spectroscopic redshift from\;\citet{fin13}. 
There is  however, a 
significant difference in the line flux, in that our grism-measured
line flux is $\sim$4 
times higher than the
measurements from the MOSFIRE spectrum.
A similar discrepancy has been seen before, in 
\citet{mas14},
where they found that the {\it HST}/WFC3 G141 grism line fluxes 
were higher  by a factor of $2-4$ compared to
the ground-based Magellan/FIRE measurements.
\ed{A similar flux comparison between HST/ACS grism and LDSS3  found a scatter of about 0.5 to 2, however
with no systematics\;\citep{xia11}.
}
In this study, while the origin of these discrepancies is not entirely clear,
possible contributing factors include
underestimation of contamination in grism spectra, 
slit-losses in ground-based spectroscopic measurements, 
underestimation of fluxes due to presence of  atmospheric  lines and much higher resolution of
ground-based spectrographs, 
and uncertainties in the absolute flux calibration.  
For \figsid\, the grism flux
calibrations seem not to be at fault since they agree with the
flux measurements from broadband images.
Whatever the cause, if 
emission line fluxes at near-IR wavelengths from  ground-based 
measurements are confirmed to be underestimates, it would reduce the
apparent strong redshift evolution in the \lya\ equivalent width  distribution. 
Firmer conclusions would benefit from a larger $z>7$ galaxy sample. Furthermore,  to minimize the
systematic errors in the \lya\ equivalent width distribution, ideally,
sample galaxies should be detected in both
\lya\ emission and the continuum, as in the case of \figsid\ (see below).

 \begin{table}[t]
  \centering
  \caption{ Properties of \figsid  }
  \begin{threeparttable}
\scriptsize {
      \begin{tabular}{llcc}			
\dtoprule

								&		Grism							&	 MOSFIRE\footnotemark[1] 		&	\\
							
  \hline
  \noalign{\vskip 0.1cm} 
  RA, DEC							&		12:36:37.913						&					&	\\
  								&		+62:18:08.60						&					&	\\
 $\lambda_{Ly\alpha}(\mu$m)			&		1.0347$\pm0.005$					&	1.0343$\pm~0.0004$			&	\\ 
$z_{Ly\alpha}$						&		7.512$\pm 0.004$ 					&	7.5078$\pm~0.0004$		&	\\
$z_{Lyman-break}$					&		7.512		 					&	$-$				&	\\
$f_{Ly\alpha}$\;($10^{-17}$\ergscm	)	&		1.06$\pm 0.19$						& 0.264$\pm~0.034$		&	\\
$f_{\lambda > 1.07\;\mu m}$\;($10^{-20}$\;\ergscm )	&	2.52$\pm 0.59$		 			& $-$				&	\\
\wlya$_{(rest)}$\;(\AA)				&		49.3$\pm 8.9$						& 7.5$\pm~1.5$		&	\\
FWHM (\AA)						&		44$\pm 9$						&	7.7$\pm~1$		&	\\
\lum ($10^{42}$\ergs)				&		7.1$\pm 1.3$						&1.77$\pm~0.36$		&	\\
$\rm\;Y_{F105W}$\;(mag)				&		26.7$\pm 0.2$						& $-$				&	\\
Lyman-break significance ($\sigma$)\footnotemark[2]	&	4.8						&  $-$				&	\\

\noalign{\vskip 0.1cm} 
\hline      
\dbottomrule
        \end{tabular}}
  \end{threeparttable}
\scriptsize {
\begin{tablenotes}
\item[]$\rm ^a$ From \citet{fin13}. 
\item[]$\rm ^b$ Based on bootstrap technique (see \S 3.5).  \\ 

%
\end{tablenotes}}

\end{table}

\subsection{Lyman break detection}
As mentioned earlier, there was no continuum break detected in the   ground-based spectroscopic observations (Finkelstein et al
2013) of  \figsid, due to shallower continuum sensitivity of MOSFIRE observations.
Using G102 grism observations of \figsid, we have detected the continuum Lyman-break, making this the second-most distant
galaxy with a spectroscopic continuum detection\;\cite[cf.][]{oes16}.
Figure~\ref{fig:stack} shows the 1D spectrum of \figsid, where the flux redward of the \lya\ line is clearly higher than the blueward  flux.
\ed{To check for any artificial break created by the G102 sensitivity (Figure~4), we performed aXesim simulations using flat input spectrum, with
a range of continuum magnitudes covering \figsid. We conclude  that the observed Lyman-break is not caused by the declining
 sensitivity function of G102.}

\ed{ The horizontal blue lines in Figure 4 show the median flux measured using the bootstrap technique where we resampled and 
replaced a flux value,  and
 remeasured the median flux value. Repeating this simulation 5000 times yields a median flux value of 
 $f_{\lambda}=0.00^{+0.49}_{-1.16}\times\;10^{-20}$\;\ergscm\, and  $f_{\lambda} =\;2.52^{+1.02}_{-0.21}\times\;10^{-20}$\;\ergscm, 
 on the blue and red side respectively.
These values yield  $\rm Y_{F105W}$ bandpass magnitude of $26.7\pm\;0.2$ mag, 
in  agreement with the measured magnitude from the imaging data,  with 
$\rm\;Y_{F105W}=26.4\;\pm\;0.2$ mag.
}

 \ed{To measure the Lyman-break significance in the presence of
   asymmetric error bars, we use the upper 
 error bar on the blue flux, and lower error bar on the red continuum
 flux. This yields a Lyman-break significance of $4.8\sigma$.
 In addition to using bootstrap technique, we directly measured the median fluxes on blue and red side. 
 This yields $f_{\lambda}=\; 0.00\;\pm\;\times\;10^{-20}$\;\ergscm\, and  $f_{\lambda}=2.52\;\pm{0.59}\;\times\;10^{-20}$\;\ergscm,
 on the blue and red side respectively. This yields Lyman-break significance of 2.7, somewhat 
 lower than  
 the bootstrap measurement, which is likely due to overestimated errors due to non-normal flux distribution.
 Thus, based on the bootstrap measurements, \figsid\ is currently the most-distant galaxy that has been spectroscopically confirmed using both the \lya\ line and the Lyman-break.
 }

\section{Summary}
Here we presented grism spectroscopy of \figsid, the first object at $z>7$ that has been spectroscopically 
confirmed using both 
the \lya\ line and the Lyman-break$-$ prior to this, $z>7$ galaxies have been confirmed
using either 
 \lya\ emission line 
or the Lyman-break detection.
\ed{
Our accurate redshift measurement based on the continuum
break detection  
demonstrates the value of FIGS and similar surveys for continuum observations.  This is crucial because as we
 probe the epoch of reionization,  we expect \lya\ emission  to attenuate, 
and therefore redshift measurements from continuum break becomes critical.
Thus, our successful identification of a galaxy in the reionization epoch  motivates planning for even more sensitive space-based grism surveys
 from upcoming missions including the {\it James Webb Space Telescope}
 and the {\it Wide Field Infrared Survey Telescope}.
}

\acknowledgements
We thank the referee for  very useful feedback that improved this manuscript.
This work is based on observations taken by the FIGS program (GO 13779)  with the NASA/ESA HST, which is operated by the Association of Universities for Research in Astronomy, Inc., under NASA contract NAS5-26555.
\ed{RAW acknowledges support from NASA JWST Interdisciplinary Scientist
grant NNX14AN10G from GSFC.}

{}

\end{document}